\documentclass[aps,prl,epsfigure,twocolumn,showpacs]{revtex4}
\usepackage{dcolumn}
\usepackage{bm}
\usepackage{graphicx}
\usepackage{amsmath}
\usepackage{latexsym}
\usepackage{amsfonts}
\usepackage{amssymb}
\usepackage{array}
\usepackage{times}

\newcommand{\ket}[1]{\left\vert#1\right\rangle}

\newcommand{\pro}[3]{\left\vert#1\rangle_{#2}\langle#3\right\vert}

\newcommand{\sprod}[2]{\left\langle#1\vert#2\right\rangle}

\newcommand{\bra}[1]{\left\langle#1\right\vert}

\begin{document}
\title{Accumulation of entanglement in a continuous variable memory}

\author{M. Paternostro$^1$, M. S. Kim$^1$ and G. M. Palma$^2$}

\affiliation{$^1$School of Mathematics and Physics, Queen's University, Belfast BT7
  1NN, United Kingdom\\
$^2$NEST-CNR (INFM) \& {D}ipartimento di Scienze Fisiche ed Astronomiche, Universita' degli studi di Palermo, via Archirafi 36, 90123, Italy }

\date{\today}

\begin{abstract}
We study the accumulation of entanglement in a memory device built out of two continuous variable (CV) systems. We address the case of a qubit mediating an indirect joint interaction between the CV systems. We show that, in striking contrast with respect to registers built out of bidimensional Hilbert spaces, entanglement superior to a single ebit can be efficiently accumulated in the memory, even though no entangled resource is used. We study the protocol in an immediately implementable setup, assessing the effects of the main imperfections.  
\end{abstract}

\maketitle

When dealing with the quantum correlations establishable within the bipartite state of two qubits, the finiteness  of the Hilbert space of the system directly sets a bound being quantified in the entanglement brought by a Bell singlet state. 
This amount is known, in literature, as an {\it ebit}. The achievement of a full ebit in a bipartite state is an important goal in many of the implementations suggested, so far, in the context of experimental quantum information processing (QIP). 
%The most emblematic example is in quantum optics, where this is routinely achieved by means of spontaneous parametric down conversion processes~\cite{spdc}. 
However, on a physical basis, there is nothing preventing us to go beyond the boundary represented by an ebit when dealing with continuous variable (CV) states.  The core of our investigation can be summarized by the following question: 'Is it possible to deposit more than a single ebit into a register?'. If this is possible, this entangled resource can be used for quantum protocols and Bell inequality tests in higher-dimensional systems~\cite{myung,teleport}. A prototypical example of a resource carrying more than an ebit can be given by looking at quantum optics, where a two-mode squeezed state carries an unbounded amount of entanglement, almost linearly dependent on the degree of squeezing $r$. This entanglement, however, is achieved via a non-linear interaction that becomes harder as $r$ increases. 

Lamata {\it et al.} have suggested a method to achieve an arbitrarily large entanglement between the motional state of two atoms by means of the projection of a two-photon state into a highly entangled subspace~\cite{lamata}. Such the projection requires a degree of non-linearity at the detection stage which is difficult to achieve with current technology. The bottleneck could be bypassed by using ancillary modes and single-photon detectors, however implying some resources overhead~\cite{lamata}.

In this paper, we discuss a scheme to beat the one-ebit bound in a CV system of two field modes (the {\it register}) with the use of an entanglement mediator. The non-linear effect in our scheme is due to the postselection of the state of the mediator after the interaction with the fields. Differently from ref.~\cite{lamata}, we do not require any projection onto entangled subspaces, the multi-ebit entanglement being the result of the re-iteration of a designed CV register-mediator interaction. Our scheme reminds the prototype for double micro-maser setups considered by Meystre {\it et al.}~\cite{meystre}, where the attention was focused on the field statistics. Phoenix and Barnett suggested the use of a single-photon cavity field as a memory for entanglement to store one ebit~\cite{phoenixbarnett}. In this and similar schemes, the one-photon subspace is used to store a single ebit. Differently, here we show that by unbounding the Hilbert space of the register, a multiple-ebit state can be prepared. 
%While our iterative protocol is suitable for the application of continuous measurement techniques~\cite{continuousmeasu}, we show that these are not effective for the task we want to address. We demonstrate that our scheme, which is implementable in various setups, can produce a multiple ebit state. 
Our scheme is also able to convert this entanglement into many entangled pairs to be used for QIP protocols.
% and a careful design of the register conditioned dynamics has to be sought. 
Our study shows the possibility of building up a quantum {\it dynamo} for useful entanglement by using an intriguing entanglement accumulation effect.

\begin{figure}[b]
%  {\bf (a)}\hskip3.5cm{\bf (b)}
\centerline{\includegraphics[width=0.3\textwidth]{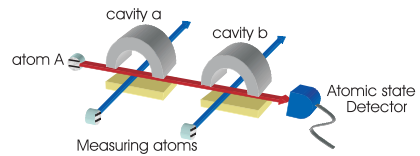}}
\caption{Setup for entanglement accumulation. The mediator $A$ interacts with cavities $a$ and $b$, prepared in independent coherent states and its state is then measured. The passage of many mediators allows for the accumulation of entanglement between the cavities. The measuring atoms allow for affirmation (and use) of the entanglement.
%We also sketch the possibility of using ancillary atomic detection for the discrimination of the amount of entanglement shared by the cavities.
}
\label{fig:fig0}
\end{figure}

{\it Protocol}.-In order to tackle our central question, we refer to a setup providing a general setting for our study. We consider two spatially separated high-quality factor cavities, each sustaining a single field mode. The decay rate of the respective photon field can therefore be neglected~\cite{commentotocome}. A flying two-level atom, which interacts with the field of each cavity, embodies the entanglement mediator. Even though we use a language typical of cavity-quantum electrodynamics (cavity-QED), our proposal is valid in those physical situations involving the interaction of a boson and a spin-like system~\cite{wilsonrae}. We sketch the suggested setup in Fig.~\ref{fig:fig0}. We consider resonant dipole-like couplings in the rotating wave approximation~\cite{knight} so that the Hamiltonian, in the interaction picture with respect to the free energy of the system, reads 
%\begin{equation}
%\label{hamiltonian}
$\hat{\cal H}=\sum_{j=a,b}\hat{\cal H}_{Aj}=\hbar\sum_{j=a,b}\lambda_j(\hat{a}_j\pro{e}{A}{g}+h.c.)$.
%\end{equation}
Here, 
% $\hat{\cal H}_{0}=E_0(\pro{e}{A}{e}+\sum_j\hat{a}^{\dag}_j\hat{a})$, 
$\hat{a}_j$ is the annihilation operator of field $j$ and $\pro{e}{A}{g}$ the spin-like raising operator of the mediator with $\ket{e}_A$ and $\ket{g}_A$ standing for its excited and ground state. The interaction-strength between each field and $A$ is $\lambda_j$. In the basis $\{\ket{e},\ket{g}\}_A$, the cavity $j$-mediator evolution after a time-interval ${t}_j$ ($\tau_{j}\!=\!\lambda_j{t}_j/\pi$ is a rescaled interaction time) is given by~\cite{phoenix}
\begin{equation}
\label{matrix}
\hat{\cal U}_{Aj}(\tau_j)=\left[\begin{matrix}\hat{\cal U}_{11}^{Aj}(\tau_j)&\hat{\cal U}_{12}^{Aj}(\tau_j)\\\hat{\cal U}_{21}^{Aj}(\tau_j)&\hat{\cal U}_{22}^{Aj}(\tau_j)\end{matrix}\right]
\end{equation}
%\begin{equation}
%\label{matrix}
%\begin{aligned}
%\hat{\cal U}_{Aj}(\tau_j)=
%\left[
%\begin{matrix}
%\cos(\pi\tau_j\sqrt{1+\hat{n}_j})&-i\hat{a}_j\frac{\sin(\pi\tau_j\sqrt{\hat{n}_j})}{\sqrt{\hat{n}_j}}\\
%i\hat{a}^\dag_j\frac{\sin(\pi\tau_j\sqrt{1+\hat{n}_j})}{\sqrt{1+\hat{n}_j}}&\cos(\pi\tau_j\sqrt{\hat{n}_j)}
%\end{matrix}\right]
%\end{aligned}
%\end{equation} 
As $[\hat{\cal H}_{Aa},\hat{\cal H}_{Ab}]\!=\!0$, the time-propagator is $\hat{\cal U}_{Ab}({\tau}_b)\hat{\cal U}_{Aa}({\tau}_a)$. By assuming minimal control over the cavity-mediator interactions, we take $\tau_{a,b}=\tau$ so to reduce the needs for a fine tuning of the mediator velocity across the cavities. 

The situation changes radically when we consider the cavities to be prepared respectively in their coherent states $\ket{\alpha}_a$ and $\ket{\beta}_{b}$. The normalized initial state of the mediator is $\ket{i}_A=c_{g}\ket{g}_A+c_e\ket{e}_A$.
% and the initial tensorial product of coherent states for the cavities is. 
The protocol consists in the projection of the mediator state at the output of cavity $b$ onto the generic single-qubit state $\ket{\theta,\varphi}_A=\cos\theta\ket{g}_A+\mbox{e}^{i\varphi}\sin\theta\ket{e}_A$. The state of the CV register resulting from this conditional dynamics is
%\begin{equation}
%\label{dinamica}
%\begin{aligned}
$\varrho^{(1)}_{ab}(\tau)\!\propto{\text{Tr}}_A\!\left\{\hat{P}_{(\theta,\phi)}\hat{\cal U}_{Ab}(\tau)\hat{\cal U}_{Aa}(\tau)\rho\,\hat{\cal U}_{Aa}^\dag(\tau)\hat{\cal U}_{Ab}^\dag(\tau)\right\}$
%\end{aligned}
%\end{equation}
with $\rho=\pro{i,\alpha,\beta}{Aab}{i,\alpha,\beta}$ the initial state of the system and $\hat{P}_{(\theta,\varphi)}=\pro{\theta,\varphi}{A}{\theta,\varphi}$ the projector describing the physical measurement of the mediator state. As a result, the register collapses into the pure state 
%$\varrho^{(1)}_{ab}(\tau)=\pro{\phi^{(1)}(\tau)}{ab}{\phi^{(1)}(\tau)}$.
%\begin{equation}
%\label{dinamica2}
$|{\phi^{(1)}(\tau)}\rangle_{ab}={\cal N}_{(\theta,\varphi)}\hat{\cal O}_{ab}(\tau)\ket{\alpha,\beta}_{ab}$
%|{\phi^{(1)}(\tau)}\rangle_{ab}\langle{\phi^{(1)}(\tau)}|={\cal N}^2_{(\theta,\varphi)}\hat{\cal O}_{ab}(\tau)\pro{\alpha,\beta}{ab}{\alpha,\beta}\hat{\cal O}^{\dag}_{ab}(\tau)
%\end{equation}
with $\hat{\cal O}_{ab}(\tau)\!=\!\mbox{}_A\!\bra{\theta,\varphi}\hat{\cal U}_{Ab}(\tau)\hat{\cal U}_{Aa}(\tau)\ket{i}_A$ and ${\cal N}_{(\theta,\varphi)}$ a normalization factor. The operator $\hat{\cal O}_{ab}$ acts just on the state of the CV register, thus showing that the postselected interaction of the mediator with each field simply result in an effective dynamics of the CV subsystems.
The form of $\hat{\cal O}_{ab}(\tau)$ can be made explicit as
% in terms of the matrix element appearing in Eq.~(\ref{matrix}) as
% by inserting the decomposition of unity $\openone_A=\pro{g}{A}{g}+\pro{e}{A}{e}$ into its expression and evaluating the resulting matrix elements. We arrive at 
$\hat{\cal O}_{ab}(\tau)=\hat{\cal A}_{b}\hat{\cal B}_{a}+\hat{\cal C}_{b}\hat{\cal D}_{a}$, where
%\begin{equation}
%\label{operators}
%\begin{aligned}
%\hat{\cal A}_{b}(\tau)=&\cos\theta\,\hat{\cal U}^{Ab}_{21}(\tau)+\mbox{e}^{-i\varphi}\sin\theta\,\hat{\cal U}^{Ab}_{11}(\tau),\\
%\hat{\cal B}_{a}(\tau)=&c_e\,\hat{\cal U}^{Aa}_{11}(\tau)+c_g\,\hat{\cal U}^{Aa}_{12}(\tau),\\
%\hat{\cal C}_{b}(\tau)=&\cos\theta\,\hat{\cal U}^{Ab}_{22}(\tau)+\mbox{e}^{-i\varphi}\sin\theta\,{\cal U}^{Ab}_{12}(\tau),\\
%\hat{\cal D}_{a}(\tau)=&c_e\hat{\cal U}^{Aa}_{21}(\tau)+c_g\hat{\cal U}^{Aa}_{22}(\tau).
%\end{aligned}
%\end{equation}
%\begin{equation}
%\label{operators}
$(\hat{\cal A}_{b},\hat{\cal C}_{b},\hat{\cal B}_{a},\hat{\cal D}_{a})^T=\hat{\cal U}^T_{Ab}\oplus\hat{\cal U}_{Aa}({\rm e}^{-i\varphi}\sin\theta,\cos\theta,c_e,c_g)^T$.
%\end{equation}
This matrix equation shows the dependence of the evolved state of the fields on the preparation of the mediator and the basis onto which its state is measured. The protocol is such that the mediator preparation (measurement) conditions the evolution of cavity $a$ ($b$). 
%Eqs.~(\ref{dinamica2}) and (\ref{operators}) are central to the study of the entanglement generation and accumulation processes.

%{\it Single-mediator passage: setting one ebit}.- 
As we mentioned, we require the measurement of the mediator state in order to bring the CV register into a pure state. This allows us to use the von Neumann entropy as a measure for the quantum correlations being established between $a$ and $b$. 
%In order to find the form of the conditioned state $\ket{\phi^{(1)}(\tau)}_{ab}$, 
An expression for arbitrary preparation of the mediator and generic $\theta,\,\varphi$ can be straightforwardly found. However, in what follows we adopt an approach directed to the simplification of the protocol. In the cavity-QED setup implicitly considered here, the preparation of the mediator state can be done by 
%populating a desired level with a classical field 
Rabi floppings between $\ket{e}_A$ and $\ket{g}_A$ using a strong classical field~\cite{harochekimble}. After the interaction with the fields, the projection onto $\ket{\theta,\varphi}_A$ requires the rotation of the mediator before the measurement of its internal state (by ionization, for instance~\cite{harochekimble}, as suggested in the caption of Fig.~\ref{fig:fig0}), therefore implying additional control over the mediator. In order to simplify the implementation, we only take $\theta=\pi/2,\,\varphi=0$, which corresponds to $\hat{P}_{(\frac{\pi}{2},0)}=\pro{e}{A}{e}$. This leads to 
\begin{equation}
\label{semplice}
%\begin{aligned}
%|{\phi^{(1)}(\tau)}\rangle_{ab}&=c_e(\hat{\cal U}^{Ab}_{11}\hat{\cal U}^{Aa}_{11}+\hat{\cal U}^{Ab}_{12}\hat{\cal U}^{Aa}_{21})\ket{\alpha,\beta}_{ab}\\
%&+c_g(\hat{\cal U}^{Ab}_{11}\hat{\cal U}^{Aa}_{12}+\hat{\cal U}^{Ab}_{12}\hat{\cal U}^{Aa}_{22})\ket{\alpha,\beta}_{ab}.
\hat{\cal O}_{ab}=c_e(\hat{\cal U}^{Ab}_{11}\hat{\cal U}^{Aa}_{11}+\hat{\cal U}^{Ab}_{12}\hat{\cal U}^{Aa}_{21})+c_g(\hat{\cal U}^{Ab}_{11}\hat{\cal U}^{Aa}_{12}+\hat{\cal U}^{Ab}_{12}\hat{\cal U}^{Aa}_{22}).
%\end{aligned}
\end{equation}
\begin{figure}[b]
  {\bf (a)}\hskip4.0cm{\bf (b)}
  \centerline{\includegraphics[width=0.2\textwidth]{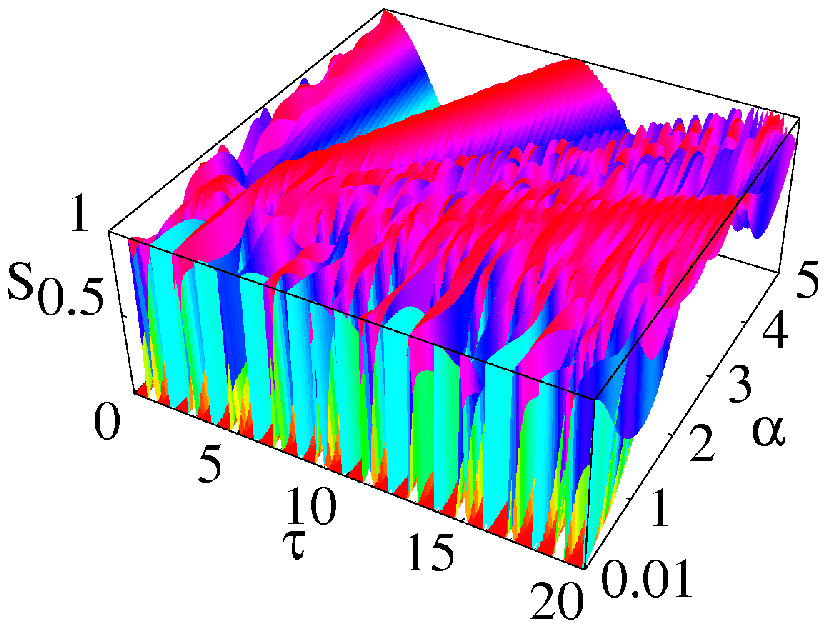}
    \hspace*{0.2cm}\includegraphics[width=0.2\textwidth]{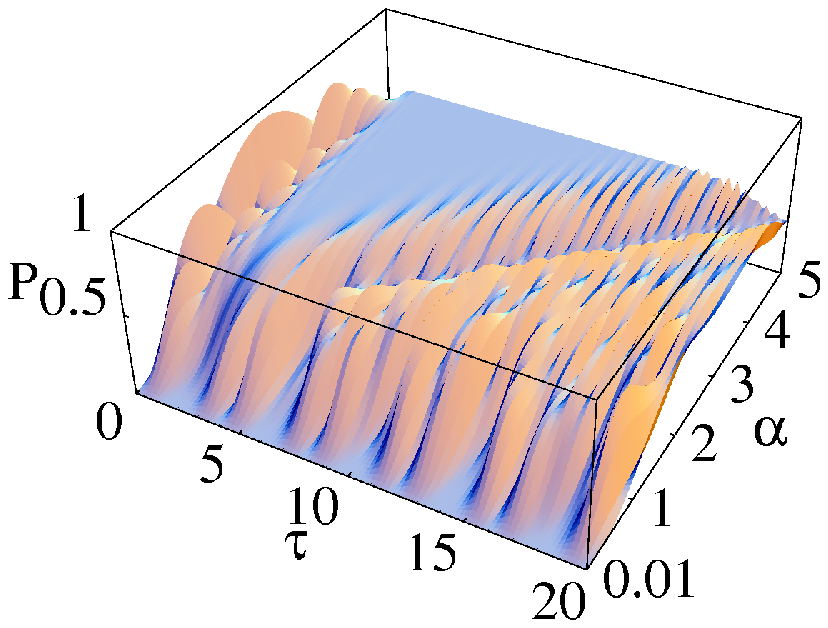}}
\caption{{\bf(a)}: Degree of entanglement between the cavity fields
  versus the interaction time $\tau$ (in unit of $\pi$) and
  the amplitude $\alpha$ of the initial coherent states. {\bf (b)}:
  Success probability of finding the atom leaving cavity $b$ in its excited
  state.}
\label{fig:fig1}
\end{figure}
A simple analysis of the symmetry properties of this operator is helpful. When highly-excited photon number states are considered, 
%the second term in Eq.~(\ref{semplice}) is such that the contributions from highly-excited photon number states will be the most considerable, 
$\hat{\cal U}^{Aj}_{11}\sim\hat{\cal U}^{Aj}_{22}$. Intuitively, one can thus expect the second term in Eq.~(\ref{semplice}) to be effective in terms of entanglement generation within the CV register, as it will be the sum of two symmetric operators. Such a symmetry does not characterize the term proportional to $c_e$. Thus, we consider the mediator to be prepared in $\ket{g}_A$.
% and, from now on, we consider explicitly just the part of $\ket{\phi(\tau)}_{ab}$ in Eq.~(\ref{semplice}) associated with $c_{g}$.
This results in $\ket{\phi^{(1)}(\tau)}_{ab}\propto\ket{\Delta}_a\ket{\Lambda}_b+\ket{\Lambda}_{a}\ket{\Gamma}_b$, where $\alpha=\beta$ has been assumed and each state vector is taken as normalized. Here, $\ket{\Lambda}_j\propto\hat{\cal U}^{Aj}_{12}\ket{\alpha}_{j},\,\ket{\Delta}_a\propto\hat{\cal U}^{Aa}_{22}\ket{\alpha}_{a}$ and $\ket{\Gamma}_b\propto\hat{\cal U}^{Ab}_{11}\ket{\alpha}_{b}$. To perform a quantitative study of the entanglement, we use the photon-number expansion $|\alpha\rangle=\sum_mC^{\alpha}_m|m\rangle$ with $C^\alpha_m=\frac{\alpha^m\mbox{e}^{-{\frac{1}{2}}\alpha^2}}{\sqrt{m!}}$ ($\alpha\!\in\!\mathbb{R}$)~\cite{barnett} so that
% with the average photon number $\bar{n}=\alpha^2$.
%\begin{equation}
%\label{fock}
%\begin{split}
%$|{\phi^{(1)}(\tau)}\rangle_{ab}\!=\!{\cal N}\!\!\sum^{\infty}_{n,m=0}[C^{\alpha}_{n}C^{\alpha+1}_{m+1}\cos(\pi\tau\sqrt{n})\sin(\pi\tau\sqrt{m+1})+C^{\alpha+1}_{n+1}C^{\alpha}_{m}\sin(\pi\tau\sqrt{n+1})\cos(\pi\tau\sqrt{m+1})]\ket{n,m}_{ab}.
$|{\phi^{(1)}(\tau)}\rangle_{ab}\!=\!{\cal N}\!\sum^{\infty}_{n,m=0}[C^{\alpha}_{n}C^{\alpha+1}_{m+1}\cos(\Theta_{n})\sin(\Theta_{m+1})+C^{\alpha+1}_{n+1}C^{\alpha}_{m}\sin(\Theta_{n+1})\cos(\Theta_{m+1})]\ket{n,m}_{ab}$
%\end{split}
%\end{equation}
with $\Theta_{p}=\pi\tau\sqrt{p}$. 
In analogy with~\cite{reciprocation}, for $\alpha^2,\tau\!\gg\!{1}$ one can approximate the Poissonian distribution characterizing a coherent state with a Gaussian distribution over $x=(n-\alpha^2)/\alpha$~\cite{barnett,reciprocation} and replace the summation with an integration over $x\in(-\infty,\infty)$. This results in $\sprod{\Gamma}{\Lambda}\propto[\sin(2\pi\tau\alpha)-{\frac{\pi\tau}{\alpha}}\cos(2\pi\tau\alpha)]\mbox{e}^{-\frac{\pi^2\tau^2}{2}}$, so that even for modest values of $\alpha$ ad $\tau$ the two states are almost orthogonal. On the other hand, there are values of $\tau$ and $\alpha$ such that the two conditions $\sprod{\Delta}{\Gamma}=1$ and $\sprod{\Delta}{\Lambda}=0$ simultaneously hold, leaving us with an equally weighted superposition of two orthogonal states: $|{\phi^{(1)}(\tau)}\rangle_{ab}$ would thus describe a CV state carrying a full ebit. We can now evaluate the entanglement in the state of the CV register after a single mediator passage by considering
%we choose to consider the exact expression , rather than the asymptotic one valid for large $\alpha$ and $\tau$. 
the entropy ${\cal S}\!=\!-\mbox{Tr}{\varrho^{(1)}_{a}}\log_2{\varrho^{(1)}_{a}}$ with 
%\begin{equation}
$\varrho^{(1)}_{a}=\mbox{Tr}_{b}(|\phi^{(1)}_{ab}(\tau)\rangle_{ab}\langle\phi^{(1)}_{ab}(\tau)|)$.
%=\sum_{m,n,n'}{\cal C}_{n,m}{\cal C}^*_{n',m}|n\rangle\langle n'|.
%\label{reduce}
%\end{equation}
% the reduced density operator for the cavity field $a$.

\begin{figure}[b]
{\bf (a)}\hskip4.0cm{\bf (b)}
\centerline{\includegraphics[width=0.2\textwidth]{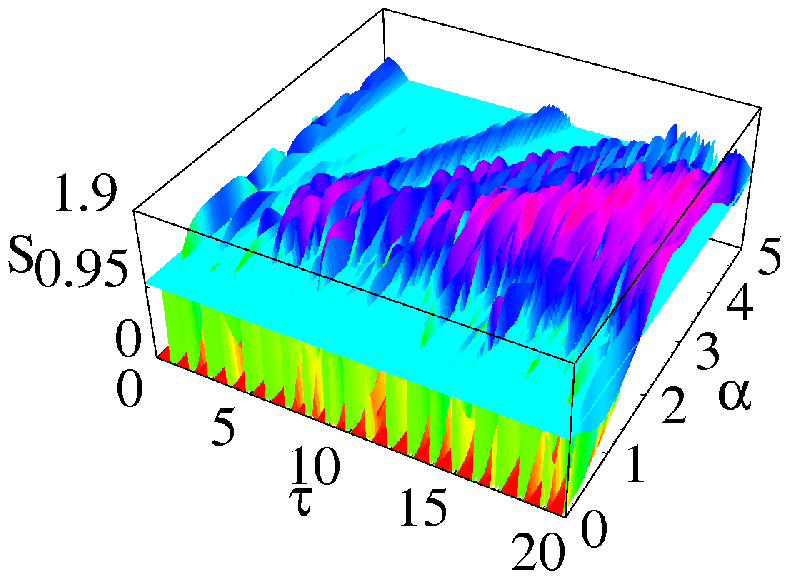}
\hspace*{0.2cm}\includegraphics[width=0.2\textwidth]{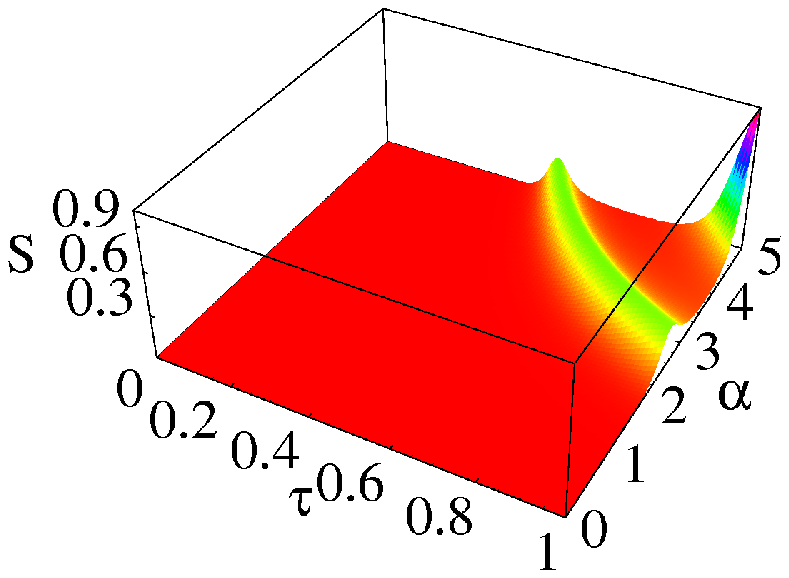}}
\caption{{\bf (a)}: Degree of entanglement between the cavity fields after the passage of the second mediator against $\tau$ and $\alpha$. A plane corresponding to ${\cal S}=1$ is shown as reference. {\bf (b)}: Degree of entanglement between the cavity fields for a negative-positive detection event plotted against $\tau$ and $\alpha$.}
%Analogous plot for the case of the entanglement reciprocation after the passage of a second pair of atoms.}
\label{fig:fig2}
\end{figure}
In Fig.1 {\bf (a)} we plot ${\cal S}$ against $\alpha$ and the interaction time $\tau$. For $\alpha=0$, we know
that ${\cal S}=0$ as no excitation can be exchanged with a mediator prepared in its ground state. However, we note that at integer values of $\tau$, this is not true even for very small amplitudes of the input coherent states~\cite{simulazione} (Fig.~\ref{fig:fig1} {\bf (a)} shows the entropy starting from $\alpha\simeq0.01$). This is very interesting but not as useful because the probability of measuring $\ket{e}_A$ for such small amplitudes is small, as shown in Fig.~\ref{fig:fig1} {\bf (b)}. When $\alpha$ increases, regions corresponding to a full ebit being shared by $a$ and $b$ are found. We identify at least two interesting regions in Fig.~\ref{fig:fig1} {\bf (a)}: The first is the long diagonal extending up to $\tau\simeq10$,
%and corresponding to short interaction time. 
the second being the triangular one involving larger $\tau$ and moderate $\alpha$'s. Both are associated with ${\cal S}=1$ and a success probability well above $45\%$.
%~\cite{commentorecipro}. 
%Physically, the reasons for the lower degree of symmetry in state (\ref{fock}) is due to the fact that a mediator prepared in its ground state can only 
%``absorb'' an excitation from the cavity field $a$.

%{\it Multi-mediator passage: entanglement accumulation}.- 
Despite the imperfect symmetry of the system at hand with respect to absorption-deposition of excitations (reflected by the fact that $\sprod{\Delta}{\Lambda}=0$ only for specific values of $\tau$ and $\alpha$), a full ebit can be conditionally created in the CV register. The central quest is, nevertheless, for larger amount of entanglement in a control-limited situation. In this context, rather than adjusting each cavity-mediator interaction (by looking for the optimal preparation of each mediator state), we assume a few identically prepared two-level systems and we postselect those cases where they are found in the respective excited states (which we call {\it positive} events). The state of the CV register resulting from the repeated application of the conditional effective operator is 
%$\hat{\cal U}^{Ab}_{11}\hat{\cal U}^{Aa}_{12}+\hat{\cal U}^{Ab}_{12}\hat{\cal U}^{Aa}_{22}$, so that
%\begin{equation}
%\label{ntimes}
%|{\phi^{(n)}}\rangle_{ab}=[\hat{\cal U}^{Ab}_{11}\hat{\cal U}^{Aa}_{12}+\hat{\cal U}^{Ab}_{12}\hat{\cal U}^{Aa}_{22}]^n\ket{\alpha,\alpha}_{ab}.
$|{\phi^{(n)}}\rangle_{ab}=\hat{\cal O}^n_{ab}\ket{\alpha,\alpha}_{ab}$.
%\end{equation}

Providing a general form for these states is increasingly difficult due to the non-commutativity of the operators appearing in $\hat{\cal O}^n_{ab}$ and one has to proceed case by case. Rather than reporting the uninformative expression for $|\phi^{(2)}\rangle_{ab}$, which is next relevant case, we show the behavior of ${\cal S}$ against $\alpha$ and $\tau$, which summarizes the salient features of a study related to entanglement accumulation. Fig.~\ref{fig:fig2} {\bf (a)} shows ${\cal S}$ for the case of a two-mediator passage, in the working conditions of limited dynamical control assumed above. If each mediator, prepared in $\ket{g}_{Ai}\,(i=1,2)$ is found in its excited state, the entanglement between $a$ and $b$ can be much larger than a single ebit. In particular, for $\alpha\ge2$ and sufficiently long interaction times, ${\cal S}$ is around $1.9$, close to the {\it optimum} of two full ebits. Interestingly, the regions of ${\cal S}<1$ are those corresponding to the non-orthogonality of $\sprod{\Gamma}{\Delta}$ in the case of a single-mediator passage. Moreover, our study shows that no accumulation effect is possible for a small-amplitude input coherent state, where the CV character of the elements of the register is not apparent. This demonstrates the distinctiveness of CV systems in the protocol suggested here.  
%\begin{figure}[b]
%  \centerline{\includegraphics[width=0.23\textwidth]{grafici/agarwalENT1}}
%\caption{Degree of entanglement between the cavity fields for a negative-positive detection event plotted versus the interaction time $\tau$ (in units of $\pi$) and the amplitude $\alpha$ of the initial coherent states. The attention is focused to short interaction times.}
%\label{fig:fig3}
%\end{figure}
%Again, the comparison with the reciprocation case is useful. The two schemes share the sub-optimality in terms of the entanglement created in the CV register: Despite both achieve two ebits of entanglement for some specific values of $\tau$ and $\alpha$, none of them is able to achieve a steady-state characterized by a two-ebit entanglement~\cite{commentonoi}. However, the higher symmetry of the reciprocation scheme allows for a result quantitatively slightly superior to the scheme here at hand. 
The entanglement accumulation protocol can be extended to additional mediators' passages, the analysis becoming more and more cumbersome due to the proliferation of terms arising from the higher-order powers of the effective dynamical operator for the CV register. Nevertheless, it is possible to see that a three-mediator passage sets entanglement around $2.5$ ebits, within the range of interaction times and $\alpha$'s considered in all the previous cases. 
Even though we do not have found evidences of ``saturation'' of the entanglement accumulation process as $n$ increases, a general feature of the scheme is that the deposition of a larger number of ebit corresponds to larger interaction times, which necessitates of better quality cavities. In the last part of this paper we address a scheme able to meet these requirements. It is interesting to notice that already for a three-mediator passage, the effect described is able to compete with the entanglement produced in a two-mode squeezed state of $r\sim{1}$~\cite{Laurat}.

We note that the {continuous measurement} theory of ref.~\cite{continuousmeasu} may be applied to the system at hand. However, in this case, the entanglement accumulation we are depicting is not effective. In the formalism of ref.~\cite{continuousmeasu}, the register interacts with a mediator for a time so short that the time-evolution operator of the system can be approximated to the second order in $\lambda$. The state of the mediator is then measured. The dynamics of the register is determined by interspersing positive with negative detection events, the latter being the cases where $\ket{g}_A$ is found. After some calculations, this results in an effective dynamics for the CV register having no entangling capabilities.
% , namely $\hat{A}=\hat{a}+\hat{b}$ and $\hat{B}(\tau)\simeq{\mbox e}^{-2R\tau(\hat{a}^\dag\hat{a}+\hat{b}^\dag\hat{b})}\mbox{e}^{-4R\tau\hat{a}\hat{b}^\dag}$, where $R=\frac{\pi^2\tau}{4}$. A dynamics in which $n$ mediators are found  in their excited state at times $\tau_{1\le{j}\le{n}}$ in an interval $(0,\tau)$ is given by $\ket{\alpha,\alpha}_{ab}\rightarrow\hat{B}(\tau-\tau_n)\hat{A}..\hat{B}(\tau_1)\ket{\alpha,\alpha}_{ab}$. As $\mbox{e}^{4R\tau_j\hat{a}\hat{b}^\dag}\hat{A}\mbox{e}^{-4R\tau_j\hat{a}\hat{b}^\dag}=(1-4R\tau_j)\hat{a}+\hat{b}$, $\mbox{e}^{2R\tau_j(\hat{a}^\dag\hat{a}+\hat{b}^\dag\hat{b})}(k\hat{a}+\hat{b})\mbox{e}^{-2R\tau_j(\hat{a}^\dag\hat{a}+\hat{b}^\dag\hat{b})}=(k\hat{a}+\hat{b})\mbox{e}^{-2R\tau_j}$ ($k\!\in\!\mathbb{R}$) and $(\mbox{e}^{-4R\tau\hat{a}\hat{b}^\dag}-\mbox{e}^{-4R\alpha\tau\hat{b}^\dag})\ket{\alpha,\alpha}_{ab}=0$, one recognizes that no entanglement can be set in the register. 
Evidence of this is given by simulating the dynamics of a two-mediator passage corresponding to a negative-positive event. No entanglement is found for short $\tau$, as shown in Fig. \ref{fig:fig2} {\bf (b)}. Only by enlarging $\tau$, quantum correlations eventually appear. However, the inclusion of a negative event in the sequence of detections limits the overall performance of the scheme. A two-mediator passage corresponds, in this case, to an entanglement slightly larger than one ebit. This study, therefore, assesses the experimentally relevant effect of a non-ideal sequence of events. The entanglement accumulation is resilient to negative events, even though it becomes less efficient.

{\it Affirmation and use of entanglement}.- In discussing suitable methods to reveal the non-classical correlations in our scheme we mention the possibility of using the recently proposed test of Bell inequalities for arbitrary numbers of measurement outcomes~\cite{myung}. Such the general framework can be adapted to the case at hand but requires computational efforts going beyond the scopes of our study. On the other hand, one can make a pragmatic use of the entanglement created in the register by arranging the passage of two auxiliary two-level system, each interacting with the respective cavity via $\hat{\cal U}_{C_jj}$ as in Eq.~(\ref{matrix}) (with $C_{j}$ the label for the auxiliary qubit crossing cavity $j=a,b$)~\cite{davidovich}. The different degree of correlation between the $C_{j}$'s, regardless of the state the cavities are left in and for different number of mediator passages can be used in order to discriminate the CV ``channels'' built out of our procedure. For instance, the state $|\phi^{(1)}\rangle_{ab}$ can be distinguished from $|{\phi^{(2)}}\rangle_{ab}$ as follows. For easiness of calculation, we refer to the case of $\tau={1}$ and $\alpha=0.8$, which is associated to ${\cal S}\simeq0.633$ ($0.994$) for $|{\phi^{(1)}}\rangle_{ab}$ ($|{\phi^{(2)}}\rangle_{ab}$). For a rescaled time of the  $C_{j}$-$j$'s interactions equal to $0.6\pi$, the corresponding entanglement between the auxiliary two-level systems is found to be ${\cal E}=0.55$ ($0.87$) with ${\cal E}$ the measure based on the negativity of partial transposition~\cite{myungDNS}. The entanglement in the $C_{a}$-$C_{b}$ system appears to be a monotonic function of the number of mediator passages through the cavities. A one-to-one correspondence between the CV entanglement and the correlations between the auxiliary qubits can be used as a tool to infer the properties of the register's state. Even though the correlations between $a$ and $b$ may be used directly (for instance for probabilistic teleportation of a multi-level system~\cite{teleport}), the entangled state of the auxiliary qubits, which  is almost pure for the case considered above, represents a valuable resource for further QIP protocols. An important feature is that, differently from $|\phi^{(1)}\rangle_{ab}$, the CV state after a multi-mediator passage can entangle sequential pairs of qubits, thus showing its role as a dynamo for entanglement.

{\it Assessment of feasibility}.- So far, we have considered the case of the fields of spatially separated cavities. This setup is implementable with state of the art technology in cavity-QED and requires steps which have all been successfully experimentally demonstrated~\cite{harochekimble}. Nevertheless, the setup can be modified so to consider just one cavity, accommodating two non degenerate, orthogonally polarized modes. With a simple modification of the Hamiltonian $\hat{\cal H}$, one can consider the interaction of the fields with a three-level mediator in a vee-configuration. We have quantitatively analyzed the modified protocol finding that multiple interactions with the mediator, prepared in an equally weighted superposition of its excited states and measured in its ground state, are able to set a multi-ebit entanglement between the independent fields.
%while multiple passages allow for the postselection of a multi-ebit entangled CV state. 
The scheme is entirely implementable and may ease an immediate realization. Indeed, a circuit-QED setup as proposed by Wallraff {\it et al}~\cite{wilsonrae} has the features of the scheme depicted here and, in addition, uses a stationary mediator (embodied by a charge qubit). The multiple passages can be simulated by ``resetting'' the state of the mediator prior to each interaction with the fields, its detection being fully implementable~\cite{wilsonrae}.

We assess the feasibility of the scheme by using values typically used in experiments. The unitary description of the process holds for $\tau$ smaller as compared to each cavity decay time $\tau_c$. In a cavity-QED setup we have $\lambda_j/\pi\simeq{50}$ KHz with $\tau_c\simeq{1}$ ms~\cite{harochekimble} ($\lambda_j/\pi\simeq{100}$ MHz with a conservative $\tau_c\simeq{160}$ ns in circuit-QED, see Wallraff {\it et al.}~\cite{wilsonrae}) allowing for tens of Rabi floppings within the coherence time of the system. Therefore, reaching the interesting regions of ${\cal S}\ge1$ is a realistic goal.
 As an experimentally relevant feature, we stress that the protocol we suggest is robust against imperfections in the setting of $\tau$, which may be due to fluctuations in $\lambda_j$ as well as in the mediator's flight speed. 
%By allowing for a mismatch between $\tau_a$ and $\tau_b$, the second following a Gaussian distribution centered in $\tau_a$ and having a width $\delta\tau$, we have verified 
In order to assess this, we fixed $\tau_a$, allowing $\tau_b$ to follow a Gaussian distribution centered at $\tau_a$ with the spread $\delta\tau$. Our calculations show that the features characterizing ${\cal S}$ persist unchanged for an experimentally motivated $\delta\tau/\tau\sim5\%$~\cite{harochekimble}. 
%For larger $\delta\tau$, the entanglement surface is slightly lowered, still preserving its salient features.

{\it Remarks}.- We have proposed a scheme for accumulating entanglement in a CV register embodied by two field modes prepared in two coherent states, demonstrating its effectiveness. We have studied the main features of the protocol and its experimental feasibility. Our proposal, which assumes minimal control over the dynamics of the mediator, is in striking contrast with the usual perspective where condensed-matter systems are assumed to store quantum properties and is sufficiently flexible to be mapped into various physical setups. Besides the purposes of this study, our scheme offers intriguing possibilities for the study of quantum channels with memory~\cite{giovmass}. 
%The analogy is clear by noticing that our protocol is designed as a {\it prepare\&{m}easure} scheme discussed in~\cite{curty}. 
The use of our proposal in simulating such the channels is a point deserving attention~\cite{commentotocome}. 

{\it Acknowledgments}.- We thank D. Murphy for valuable help in programming and C. Di Franco, J. Lee and W. Son for discussions. We acknowledge support from the Leverhulme Trust (ECF/40157) and the UK EPSRC.


\begin{thebibliography}{99}
\bibitem{simulazione} The state associated with $\tau\!=\!1$ and $\alpha=0.01$ is $0.707(\ket{01}-\ket{10})_{ab}+0.003(\ket{02}-\ket{20})_{ab}-0.009\ket{11}_{ab}+{\cal O}(10^{-5})$, whose entropy is ${\cal S}\sim0.999$.
\bibitem{davidovich} This configuration may remind of the scheme for atomic teleportation by L. Davidovich {\it et al.}, \pra {\bf 50}, R895 (1994) where, however, only the one-photon subspace was considered.
\end{thebibliography}
\end{document}